\documentstyle[12pt]{article}
\begin{document}

\title{Elementary excitations of a Higgs-Yukawa system}

\author{
E. R. Takano Natti\\
Pontif\'{\i}cia Universidade Cat\'olica do Paran\'a\\
Rua J\'oquei Clube, 458, 86067-000, Londrina, PR, Brazil\\
(erica.natti@pucpr.br) \\ \\
A. F. R. de Toledo Piza\\
Instituto de F\'{\i}sica, Universidade de S\~ao Paulo \\
Caixa Postal 66318, Cep 05315-970, S\~ao Paulo, SP, Brazil\\
(piza@fma.if.usp.br) \\ \\
P. L. Natti\\
Departamento de Matem\'atica,
Universidade Estadual de Londrina\\
86051-990 Londrina, PR, Brazil\\
(plnatti@uel.br) \\ \\
Chi-Yong Lin\\
Department of Physics, National Dong Hwa University, \\
Hualien, Taiwan, Republic of China\\
(lcyong@mail.ndhu.edu.tw)}

\maketitle
\begin{abstract}

  This work investigates the physics of elementary excitations for the
  so-called relativistic quantum scalar plasma system, also known as
  the Higgs-Yukawa system. Following the Nemes-Piza-Kerman-Lin
  many-body procedure, the Random-Phase Approximation (RPA) equations
  were obtained for this model by linearizing the Time-Dependent
  Hartree-Fock-Bogoliubov equations of motion around equilibrium. The
  resulting equations have a closed solution, from which the spectrum
  of excitation modes are studied. We show that the RPA oscillatory
  modes give the one-boson and two-fermion states of the theory.  The
  results indicate the existence of bound states in certain regions in
  the phase diagram.  Applying these results to recent LHC
  observations concerning the mass of the Higgs boson, we determine
  limits for the intensity of the coupling constant $g$ of the
  Higgs-Yukawa model, in the RPA mean-field approximation, for three
  decay channels of the Higgs boson.  Finally, we verify that, within
  our approximations, only Higgs bosons with masses larger than $190$
  GeV/$c^2$ can decay into top quarks.


\vskip 0.5cm


\end{abstract}

\section{Introduction}

Recent years have witnessed substantial progress towards understanding
the non-equilibrium time evolution of quantum fields. Results have
been obtained that proved important in several applications. Examples
are found in cosmology, such as the description of quantum-field
expectation values in the early universe and subsequent hot stage (big
bang) \cite{JB04,GHKL10}, in high energy particle physics, such as the
description of the dynamics in heavy-ion collision experiments, which
seeks to establish experimental signatures for the non-equilibrium
evolution of the quark-gluon system and chiral phase transition
\cite{CCMS01,MS05,FSKR07}, and in complex many-body quantum systems, such
as the description of the dynamics of the Bose Einstein condensates
\cite{LPTPLH06,GLG10}, among other applications \cite{EHKKMWW10}.

In a previous publication we have developed a framework to investigate
the initial-value problem in the context of interacting fermion-scalar
field theories \cite{TNa98}. This framework had earlier been developed in
the context of the non-relativistic nuclear many-body dynamics by
Nemes and Toledo Piza \cite{TP82}.  The method led to a set of
self-consistent time-dependent equations for the expectation values of
linear and bilinear forms of field operators.  These dynamical
equations acquire a kinetic type structure in which the lowest-order
approximation corresponds to the usual Gaussian mean-field
approximation.  As an application, a zero-order calculation was
implemented within the simplest context of a relativistic quantum
scalar plasma system, nowadays also called the Higgs-Yukawa system.
The usual renormalization prescription was shown to be also
applicable to this non-perturbative calculation.  In particular, a
finite expression for the energy density was obtained, the numerical
results suggesting that the system always has a single stable minimum.

Here we report an application of the renormalized nonlinear dynamical
equations obtained in our previous article \cite{TNa98} and
follow Kerman and Lin \cite{KL95,KL98} to investigate the
near-equilibrium dynamics around the stationary solution of a
Higgs-Yukawa system. We will show that one-boson and two-fermion
physics can be studied in the linear approximation of the
mean-field equations.  We will solve those equations in
closed form and find the scattering amplitude, as well as the
conditions allowing a two-fermion bound state.

The motivation for this work is the recent observation of a possible
Higgs boson around 125 Gev/$c^2$ \cite{Higgs1}-\cite{Higgs3}.  Since
the early 1990's, the Higgs-Yukawa model has been used to better understand
the fermion mass generation via the Higgs mechanism. Recently, the
Higgs-Yukawa model has been used to impose limits on the Higgs
mass and on the intensity of the Higgs-Yukawa coupling.  It has also been
used to study the consequences of heavy
extra-generation fermions (mass $>$ 600/$c^2$ GeV).  An important
consequence of a fourth fermion generation is the possibility of
formation of bound states that can replace the role of the Higgs
boson. Recently, these issues have been intensively studied
on the lattice \cite{Higgs2012} or perturbatively\textemdash the
$1/N$-expansion \cite{Higgs2011}, for example.
Our work carries out a non-perturbative calculation of
the ground state (vacuum) for an interacting Higgs-Yukawa system.

For clarity and notational purposes, a few key equations from
\cite{TNa98} are repeated here.  The dynamics of the relativistic
quantum scalar plasma model, or Higgs-Yukawa model, is governed by the
following lagrangian density \cite{Kalman}-\cite{Fraga}:
\begin{eqnarray}
\label{1abc}
{\cal L}&=&\bar\psi
(i \gamma.{\partial}-m)\psi+
g\bar\psi\phi\psi \nonumber\\ \nonumber\\
&&+\frac{1}{8\pi}
\left[(\partial\phi)^2-
\mu^2\phi^2\right]-{\cal L}_{\mbox {\tiny C}}\;\;.
\end{eqnarray}

The model Hamiltonian is given by the equality
\begin{eqnarray}
\label{1}
H&=&\int_{\bf x} {\cal H}\;,\nonumber\\ \nonumber\\
{\cal H}&=&-\bar\psi
(i\vec\gamma.{\vec\partial}-m)\psi-
g\bar\psi\phi\psi \nonumber\\ \nonumber\\
&&+\frac{1}{8\pi}
\left[(4\pi)^2 \Pi^2+|\partial\phi|^2+
\mu^2\phi^2\right]+{\cal H}_{\mbox {\tiny C}}\;\;,
\end{eqnarray}
with the shorthand $\int_{\bf x}=\int d^{3}x$.

In the Heisenberg picture, $\phi(x)$ and $\Pi(x)$ are scalar spin-$0$ fields
expanded as
\begin{eqnarray}
\label{2}
\phi({\bf x},t) &=&\sum_{\bf p} \frac{1}{(2Vp_{0})^{1/2}} \left[b_{\bf
p}(t)e^{i{\bf p}.{\bf x}}+b_{\bf p}^{\dag}(t) e^{-i{\bf p}.{\bf
x}}\right]\;\;\nonumber\\ \\
\Pi({\bf x},t) &=&i\sum_{\bf p} \left(
\frac{Vp_{0}}{2}\right)^{1/2} \left[b_{\bf p}^{\dag}(t)e^{-i{\bf
p}.{\bf x}} -b_{\bf p}(t)e^{i{\bf p}.{\bf x}}\right]\;\;,\nonumber
\end{eqnarray}
where $b_{\bf p}^{\dag}(t)$ and $b_{\bf p}(t)$ are boson creation
and annihilation operators, $p_{0}=\sqrt{{\bf p}^2 +\Omega^2}\;$,
$\Omega$ is the mass parameter of the bosonic fields,
$\;p\:x=p_{0}\:t-{\bf p}.{\bf x}$, and $\psi(x)$ and
$\bar\psi(x)$ are fermionic spin-$1/2$ fields,
\begin{eqnarray}
\label{3}
\psi({\bf x},t)&=&\sum_{{\bf
k},s}\left(\frac{M}{k_{0}}\right)^{1/2}\frac{1}{\sqrt V}
\left[u_{1}({\bf k},s)a_{{\bf k},s}^{(1)}(t)e^{i {\bf k}.{\bf x}}
+u_{2}({\bf k},s){a_{{\bf k},s}^{(2)}}^{\dag}(t)
e^{-i {\bf k}.{\bf x}}\right]
\nonumber\\ \\
\bar\psi({\bf x},t)&=&\sum_{{\bf
k},s}\left(\frac{M}{k_{0}}\right)^{1/2}\frac{1}{\sqrt V}\left[\bar
u_{1}({\bf k},s){a_{{\bf k},s}^{(1)}}^{\dag}(t)e^{-i {\bf k}.{\bf x}}+
\bar u_{2}({\bf k},s)a_{{\bf k},s}^{(2)}(t)
e^{i {\bf k}.{\bf x}}\right]\;\;,
\nonumber
\end{eqnarray}
where ${a_{{\bf k},s}^{(1)}}^{\dag}(t)$ and
$a_{{\bf k},s}^{(1)}(t)$ [${a_{{\bf k},s}^{(2)}}^{\dag}(t)$ and
$a_{{\bf k},s}^{(2)}(t)$] are fermion creation and
annihilation operators associated with the positive- (negative-) energy
solutions $u_{1}({\bf k},s)$ [$u_{2}({\bf k},s)$] of Dirac's
equation. Likewise, $k_{0}=\sqrt{{\bf k}^2 +M^2}\;$,
$M$ is the mass parameter of the fermionic fields and
$\;k\:x=k_{0}\:t-{\bf k}.{\bf x}$ .

In Eq.~(\ref{1}), the parameters $m$ and $\mu$ are the masses
of the fermion and of the scalar particles, respectively, and $g$ is the coupling constant.
The last term on the-right hand side, which encompasses the counterterms necessary to
remove the later-occuring infinities, is given by the expression
\begin{equation}
\label {27}
4\pi {\cal H}_{\mbox {\tiny C}} = \frac{A}{1!}\phi +
\frac{\delta \mu^{2}}
{2!}\phi^{2} + \frac{C}{3!}\phi^{3} + \frac{D}{4!}\phi^{4} -
\frac{Z}{2} (\partial\phi)^2 \;\;,
\end{equation}
where the coefficients $A$, $\delta \mu^{2}$, $C$ and $D$
are infinite constants, to be defined later. To study the dynamical
RPA regime, we have to introduce the wave-function renormalization
constant $Z$ \cite{AH88}.

To deal with condensate and pairing dynamics of the scalar and
fermionic fields, we first define the unitary
Bogoliubov transformation for the bosonic sector as follows \cite{RS80}:
\begin{equation}
\label{4}
\left[
\begin{array}{c}
d_{\bf p}(t)\\
        \\
d_{-\bf p}^{\dag}(t)
\end{array}
\right]=\left[
\begin{array}{cc}
\cosh \kappa_{\bf p}+i\frac{\eta_{\bf p}}{2}  &
-\sinh \kappa_{\bf p}+i\frac{\eta_{\bf p}}{2}\\ \\
-\sinh \kappa_{\bf p}-i\frac{\eta_{\bf p}}{2} &
\cosh \kappa_{\bf p}-i\frac{\eta_{\bf p}}{2}
\end{array}
\right]
\left[
\begin{array}{c}
\beta_{\bf p}(t)  \\
      \\
\beta_{-\bf p}^{\dag}(t)
\end{array}
\right]\;\;,
\end{equation}
where $d_{\bf p}$ is the shift boson operator defined by the expression

\begin{equation}
\label{4a}
d_{\bf p}(t)\equiv b_{\bf p}(t)- B_{\bf p}(t)\delta_{{\bf
p},0}\;\;\;\;{\mbox{with}}\;\;\;\;
B_{\bf p}(t) \equiv\langle b_{\bf p}(t) \rangle = Tr_{\mbox{\tiny
BF}}\;\left[b_{\bf p}(t){\cal F}\right]\;\;.
\end{equation}
Here,  $\cal F$ is the unitary many-body density operator \cite{TNa98}
describing the system state, and the symbol $Tr_{\mbox{\tiny BF}}$
denotes a trace over both bosonic and fermionic variables. Partial
traces over bosonic or fermionic variables will be denoted
$Tr_{\mbox{\tiny B}}$ and $Tr_{\mbox{\tiny F}}$, respectively.
For simplicity, we restrict our treatment of the fermionic sector to
the Nambu transformation \cite{NTP96}, parameterized in
the following form, which incorporates the unitarity constraints:
\begin{equation}
\label{5}
\left[\!
\begin{array}{c}
      a_{{\bf k},s}^{(1)}    \\
                   \\
      a_{{\bf k},s}^{(2)}    \\
                   \\
{a_{-{\bf k},s}^{(1)}}^{\dag}\\
                   \\
{a_{-{\bf k},s}^{(2)}}^{\dag}
\end{array}
\!\right]\!=\!\left[\!
\begin{array}{cccc}
\cos\varphi_{\bf{k}} &      0       &      0       &
          -e^{-i\gamma_{\bf{k}}}\sin\varphi_{\bf{k}} \\
          &              &              &              \\
  0       & \cos\varphi_{\bf{k}} & e^{-i\gamma_{\bf{k}}}
\sin\varphi_{\bf{k}} &     0        \\
          &              &              &              \\
  0       & -e^{i\gamma_{\bf{k}}}\sin\varphi_{\bf{k}}
          &\cos\varphi_{\bf{k}} &     0        \\
          &              &              &              \\
e^{i\gamma_{\bf{k}}}\sin\varphi_{\bf{k}} &      0       &      0
          & \cos\varphi_{\bf{k}}
\end{array}
\!\right]\left[
\!\begin{array}{c}
    \alpha_{{\bf k},s}^{(1)}    \\
                         \\
    \alpha_{{\bf k},s}^{(2)}    \\
                         \\
{\alpha_{-{\bf k},s}^{(1)}}^{\dag}\\
                         \\
{\alpha_{-{\bf k},s}^{(2)}}^{\dag}
\end{array}
\!\right]\;\;.
\end{equation}

The procedure adopted in \cite{TNa98} to obtain the equations of motion for
the Nambu-Bogoliubov parameters $\varphi_{\bf k}(t)$ , $\gamma_{\bf
  k}(t)$, $\eta_{\bf p}(t)$, $\kappa_{\bf p}(t)$, for the
quasi-particle occupation numbers $\nu^{(\lambda)}_{{\bf k},s}
=\langle{\alpha^{(\lambda)}_{{\bf k},s}}^{\dag}
\alpha^{(\lambda)}_{{\bf k},s}\rangle$, $\;\nu_{\bf p}
=\langle\beta_{\bf p}^{\dag}\beta_{\bf p}\;\rangle$, and for the
condensates $\langle\phi\rangle$ and $\langle\Pi\rangle$ had been
developed earlier in the context of the non-relativistic nuclear
many-body dynamics by Nemes and Toledo Piza \cite{TP82}.  That approach
follows the line of thought of a time-dependent projection technique
proposed by Willis and Picard \cite{WP74} in the context of the master
equation for coupled systems.  The method consists, essentially, of
writing the correlation information of the full density of the system
$\cal F$ in terms of a memory kernel acting on the uncorrelated
density ${\cal F}_{0}$, with the help of a time-dependent projector.
At this point, a systematic mean-field expansion for two-point
correlations can be performed \cite{TP82}.  The lowest order
corresponds to the results of the usual Gaussian approximation
\cite{NTP96,LTP92}.  The higher orders describe the dynamical
correlation effects between the subsystems and are expressed by means
of suitable memorial integrals added to the mean-field dynamical
equations.  The resulting equations acquire the structure of
kinetic equations, with the memory integrals playing the role of collisional
dynamics terms. This systematic expansion scheme for memory effects,
in which the mean energy is conserved to all orders
\cite{LW60,BA62,BFN88,IKV99}, was implemented, for example, for the
Jaynes-Cummings system \cite{ETP97}.

In this context, to study the near-equilibrium dynamics
around the stationary solution of the Higgs-Yukawa system,
Ref.~\cite{TNa98} focused on a selected
set of Gaussian observables, which are related to the expectation
values of linear, $\phi(x)$, $\Pi(x)$, and bilinear forms of field operators,
such as $\phi(x)\phi(x)$ , $\bar\psi(x)\psi(x)$ ,
$\psi(x)\psi(x)$, etc.. The time evolution of such
quantities obeys the Heisenberg equation of motion
\begin{equation}
\label{31a}
i\langle \dot{\cal O} \rangle
=Tr_{\mbox{\tiny BF}}[{\cal O},H] {\cal F}\;\;,
\end{equation}
where ${\cal O}$ can be $\phi(x)$, $\phi(x)\phi(x)$,
$\bar\psi(x)\psi(x)$, etc., and $\cal F$ is the state of the system,
which is assumed to be spatially uniform, in the Heisenberg
picture. As an approximation, we replace the full density $\cal F$ by
a truncated ansatz ${\cal F}_{0}(t)$, whose trace is also unitary, and
which implements the double mean-field approximation \cite{TNa98}.  By
construction, ${\cal F}_{0}$ is written as the most general Hermitian
Gaussian functional of the field operators consistent with the assumed
uniformity of the system. It will thus be written as the exponential
of a general quadratic form in the field operators, which can be
reduced to diagonal form by a suitable canonical transformation. In
this way, ${\cal F}_{0}$ reproduces the corresponding ${\cal F}$
averages for linear or bilinear field operators \cite{NTP96}.  In
particular, we have used a formulation appropriate for the many-body
problem, where ${\cal F}_{0}$ is written in the momentum basis as
\cite{NTP96,LTP92,ETP97}
\begin{eqnarray}
\label{31b}
{\cal F}_{0}
&=&{\cal F}^{\mbox{\tiny B}}_{0} {\cal F}^{\mbox{\tiny F}}_{0}
\nonumber\\
\label{32a}
{\cal F}^{\mbox{\tiny B}}_{0}
&=&\prod_{\bf p}\left[\frac {1} {1+\nu_{\bf
p}}\left(\frac {\nu_{\bf p}}{1+\nu_{\bf p}}\right) ^{\beta_{\bf
p}^{\dag}\beta_{\bf p}} \right]\\ \nonumber\\
\label{32b}
{\cal F}^{\mbox{\tiny F}}_{0}
&=&\prod_{{\bf k},s,\lambda} \left[\nu_{{\bf
k},s}^{(\lambda)}{\alpha^{(\lambda)}_{{\bf k},s}}^{\dag}
\alpha^{(\lambda)}_{{\bf k},s}+(1-\nu_{{\bf k},s}^{(\lambda)})
\alpha_{{\bf k},s}^{(\lambda)}{\alpha^{(\lambda)}_{{\bf
k},s}}^{\dag} \right] \;\;,
\end{eqnarray}

\noindent
where $\alpha$ ($\alpha^{\dag}$) and $\beta$
($\beta^{\dag}$) stand for Nambu-Bogoliubov quasi-particle annihilation
(creation) operators for fermions and bosons, respectively,
$\nu^{(\lambda)}_{{\bf k},s}$
$\left(\;\nu_{\bf p}\right)$
are the quasi-fermion (quasi-boson) occupation numbers, and
$\lambda=1$ $\left(\lambda=2\right)$ is associated with the positive-
(negative-) energy solutions.

From Eqs.~(\ref{31a})-(\ref{32b}) we can directly obtain the equation of motion for the
occupation numbers, for the Gaussian variables, now represented by the
Nambu-Bogoliubov parameters, and for the condensates. The following expressions result:
\begin{eqnarray}
\label{41}
\dot\nu_{{\bf p}}&=&\dot\nu^{(1)}_{{\bf k},s}=\dot\nu_{{\bf k},s}^{(2)}=0\\\nonumber\\
\label{41a}
\dot \varphi_{\bf k}&=& \frac{|{\bf k}|}{k_{0}}(M-\bar{m})
\sin\gamma_{{\bf k}}\\\nonumber\\
\label{42}
\sin2\varphi_{{\bf k}}\dot\gamma_{{\bf k}}&=&
\frac{2({\bf k}^2+\bar{m}M)}{k_{0}}\sin2\varphi_{{\bf k}}\nonumber\\\nonumber\\
&+&\frac{2(M-\bar{m})}{k_{0}} {|{\bf k}|}
\cos2\varphi_{{\bf k}}\cos\gamma_{{\bf k}}\\\nonumber\\
\label{42s}
\dot\eta_{\rm p}e^{-\kappa_{\rm p}}&=&({\bf p}^2+\Omega^2)^{1/2}
\left[4\pi e^{2\kappa_{\rm p}}-\frac{1}{4\pi}\frac{({\bf p}^2+\mu^2)}
{({\bf p}^2+\Omega^2)}e^{-2\kappa_{\rm p}}\right]\;\;\\\nonumber\\
\label{42r}
\dot\kappa_{\rm p}&=&-4\pi({\bf p}^2+\Omega^2)^{1/2}
\eta_{\rm p}e^{\kappa_{\rm p}}\\\nonumber\\
\label{43}
\langle\dot\phi\rangle&=&\frac{4\pi}{(1+Z)}\langle\Pi\rangle\\\nonumber\\
\label{44}
\langle\dot\Pi\rangle
&=&-\frac{1}{4\pi}\left[A+\frac{C}{2}G(\Omega)\right]\nonumber\\\nonumber\\
&-&\frac{1}{4\pi}\left[\mu^2+\delta\mu^2+\frac{D}{2}G(\Omega)\right]
\langle\phi\rangle-\frac{C}{8\pi}\langle\phi\rangle^2\nonumber\\\nonumber\\
&-&\frac{D}{24\pi}\langle\phi\rangle^3
-g\sum_{s} \int_{\bf k} \frac{1}{k_{0}}[M\cos 2\varphi_{{\bf k}}
\nonumber\\\nonumber\\
&+&{|{\bf k}|} \sin 2\varphi_{{\bf k}}\cos\gamma_{{\bf k}}]
(1-\nu_{{\bf k},s}^{(1)}-\nu_{{\bf k},s}^{(2)})\;\;,
\end{eqnarray}
\vskip 0.3cm
\noindent
where we have introduced the notation
\begin{equation}
\label{45}
G(\Omega) =
\int_{\bf p} \frac{1+2\nu_{\rm p}}{2\sqrt{{\bf p}^2+{\Omega}^2}} \;\;.
\end{equation}
In Eqs.~(\ref{41})-(\ref{45}) the quantities $M$ and $\Omega$ are the mass
parameters of the fermionic $\psi$ and bosonic $\phi$ fields of the
Hamiltonian~(\ref{1}), while $\bar m\equiv m-g\langle\phi\rangle$
stands for the effective mass of fermion particle \cite{TNa98}.

Another physical quantity of interest is the mean-field energy density
of the system,
\begin{eqnarray}
\label{123a}
&&\frac{\langle H\rangle}{V} = \frac{1}{V}Tr H{\cal F}_{0}
\nonumber\\\nonumber\\
&&=-\sum_{s}\int_{\bf k} \left[\frac{({\bf k}^2+\bar m{M})}{k_{0}}
\cos 2\varphi_{{\bf k}}+\frac{(\bar m-M)}{k_{0}} {|{\bf k}|}
\sin 2\varphi_{\bf k} \cos\gamma_{\bf k}\right](1-\nu_{{\bf k},s}^{(1)}-
\nu_{{\bf k},s}^{(2)})\nonumber\\\nonumber\\
&&+\frac{1}{8\pi}\left[\langle\Pi\rangle^2+
\mu^2\langle\phi\rangle^2\right]
+\frac{1}{4\pi}\left[A+\frac{C}{2}G(\Omega)\right]
\langle\phi\rangle
+\frac{1}{8\pi}\left[\delta\mu^2+\frac{D}{2}G(\Omega)\right]\langle\phi\rangle^2
\nonumber\\\nonumber\\
&&+\frac{C}{24\pi}\langle\phi\rangle^3
+\frac{D}{96\pi}\langle\phi\rangle^4
+\frac{1}{8\pi}\left[\mu^2+\delta\mu^2\right]G(\Omega)
+\frac{D}{32\pi}G^2(\Omega)\;\;.
\end{eqnarray}

The above equations describe the real-time evolution of the
Higgs-Yukawa system in the double Gaussian mean-field
approximation. The results obtained in (\ref{41})-(\ref{123a}),
as discussed in \cite{TNa98}, are consistent with those in the literature,
obtained via different approaches, in particular with those obtained in
Ref.~\cite{AH84} on the basis of a Vlasov-Hartree approximation.

Reference~\cite{TNa98} showed, in detail, that the usual form of
renormalization \cite{AH84} is applicable to the non-perturbative
procedure described in that paper.  A simple numerical calculation has
also shown that the system always has a single stable minimum,
although, as it has been suggested \cite{TNa98}, additional
investigation is necessary concerning oscillatory modes.  The standard
approach to this question uses the RPA analysis, the resulting
eigenvalues giving an indication of stability \cite{KL95,KL98,AH88}.

Finally, we note that dynamical correlation corrections can in
principle be systematically added to the double-Gaussian mean-field
calculations with the help of a projection technique discussed in
\cite{TP82,BFN88,ETP97}.  The occupation numbers are then no longer
constant, a modification that affects the effective dynamics of the
Gaussian observables.  The framework presented in this paper also
serves as groundwork for finite-density and finite-temperature
discussions \cite{TTP97}.  In particular, a finite matter-density
calculation beyond the mean-field approximation allows one to study
such collisional observables as the transport coefficients.  The
extension of this procedure to nonuniform systems is straightforward,
albeit long.  In this case, the spatial dependence of the field are
expanded in natural orbitals of the extended one-body density.  A more
general Bogoliubov transformation \cite{RS80} would relate these
orbitals to a momentum expansion.

The results in Eqs.~(\ref{41})-(\ref{44}) are nonlinear time-dependent
field equations. A closed solution is not easily constructed.  Here we
consider those equations in the small oscillation regime and find a
closed solution offering insight into divers properties of the theory.

The paper is structured as follows. In Sec.~\ref{sec:2}, the
RPA equations are derived for this model by considering
near-equilibrium dynamics around the stationary solutions obtained in
Ref.~\cite{TNa98}.  Section~\ref{sec:3} finds analytical solutions for
the RPA equations by using a well-know procedure from the scattering
theory.  Section~\ref{sec:4} discusses renormalization within the
context of scattering amplitudes and discusses the existence of bound
state solutions. Finally, we apply these results to find limits
to the intensity of the coupling constant $g$ of the Higgs-Yukawa
model, in the RPA mean-field approximation, for three decay channels
of the Higgs boson.  Section~\ref{sec:5} presents our conclusions.
\label{sec:1}
\section{Near equilibrium dynamics}
\label{sec:2}
The energy density~(\ref{123a}) is a function of the Nambu-Bogoliubov
parameters $\varphi_{\bf k}(t)$ , $\gamma_{\bf k}(t)$, $\eta_{\bf
  p}(t)$, $\kappa_{\bf p}(t)$, of the quasi-particle occupation
numbers $\nu^{(\lambda)}_{{\bf k},s} =\langle{\alpha^{(\lambda)}_{{\bf
      k},s}}^{\dag} \alpha^{(\lambda)}_{{\bf k},s}\rangle$,
$\;\nu_{\bf p} =\langle\beta_{\bf p}^{\dag}\beta_{\bf p}\;\rangle$,
and of the condensates $\langle\phi\rangle$ and $\langle\Pi\rangle$.
The minimum in Eq.~(\ref{123a}) corresponds to the ground state of the
system.  The small amplitude motion around the minimum is obtained by
linearization of the Gaussian motion equations (\ref{41})-(\ref{44}),
yielding a set of harmonic oscillators \cite{KK76}. The eigenvalues
and the normal modes of these small oscillations are the RPA
solutions. Physically, the RPA solutions are seen as the energy and
the wave functions of quantum particles. This section derives the RPA
equations of the model, whose solutions are discussed in the Sec.~\ref{sec:3}.

First, we have to consider the stationary problem \cite{TNa98}.
We only have to recall Eqs.~(\ref{41})-(\ref{44}) to see that
$\dot\gamma_{\bf k}=\dot\varphi_{\bf k}= \dot\kappa_{\bf p}
=\dot\eta_{\bf p}=\langle\dot\phi\rangle=\langle\dot\Pi\rangle=0$
under stationary conditions. Reference~\cite{TNa98} discussed the renormalization conditions and the solutions
for this set of stationary equations in detail.
In particular, for the renormalization coefficients of ${\cal H}_{c}$,
the following self-consistency renormalization prescription was
chosen \cite{TNa98,AH88}:
\begin{eqnarray}
\label{94}
D&=&\pm48\pi g^4 L(m)\;\\\nonumber\\
\label{95}
\delta\mu^2&=&\mp24\pi g^4L(m)G(\mu)\mp16\pi
g^2G(0)\pm24\pi m^2g^2 L(m)\;\\\nonumber\\
\label{96}
C&=&\mp48\pi mg^3L(m)\;\\\nonumber\\
\label{97}
A&=&\pm24\pi mg^3L(m)G(\mu)\pm16\pi mgG(m)\;
\end{eqnarray}
with
\begin{equation}
\label{98}
L(m) \equiv \int_{\bf k} \frac{1}{2{\bf k}^2 ({\bf k}^2+m^2)^{1/2}}\;\;,
\end{equation}
where $M=m$, without loss of generality \cite{TNa98}.

The substitution of such counterterms in the stationary equations
yields the appropriate cancelations, which makes the equations finite,
except for the combination of the type $ L(m)[G(\mu)-G(\Omega)]$.  Since
$\Omega$ is an arbitrary expansion mass parameter, one can remove this
divergence by setting $\Omega=\mu$ \cite{TNa98}.  The resulting
finite stationary equations for the system can be regrouped as follows:
\begin{eqnarray}
\label{35}
&&\sin\gamma_{\bf k}|_{\mbox{\tiny eq}}=0\\\nonumber\\
\label{36}
&&\cot 2\varphi_{\bf k}|_{\mbox{\tiny eq}}=-\frac{({\bf k}^{2}+{\bar m}m)}
{|{\bf k}|(m-{\bar m})}\\\nonumber\\
\label{36p}
&&\eta_{\rm p}|_{\mbox{\tiny eq}}=0\\\nonumber\\
\label{36q}
&&\kappa_{\bf p}|_{\mbox{\tiny eq}}=0\;\;\;{\mbox{with}}\;\;\;\Omega=\mu\\\nonumber\\
\label{39}
&&\langle\Pi\rangle|_{\mbox{\tiny eq}}=0\\\nonumber\\
\label{40}
&&\frac{\pi}{2}{\mu}^2\langle\phi\rangle|_{\mbox{\tiny eq}}
-g{\bar m}^3\left[\ln\left(\frac{\bar m}{m}\right)+\frac{1}{2}\right]=0\;\;.
\end{eqnarray}

Equations (\ref{35})-(\ref{40}) can be numerically solved for any
given $\mu$ and $g$, in units of $m$, as shown by Ref.~\cite{TNa98}.

To obtain the near-equilibrium dynamics (RPA regime), we examine the fluctuations around
the stationary solution, namely,
\begin{eqnarray}
\label{48}
\varphi_{\bf k}&=&\varphi_{\bf k}|_{\mbox{\tiny eq}}
+\delta\varphi_{\bf k}
\nonumber\\\nonumber\\
\gamma_{\bf k}&=&\gamma_{\bf k}|_{\mbox{\tiny eq}}
+\delta\gamma_{\bf k}\nonumber\\\nonumber\\
\eta_{\bf p}&=&\eta_{\bf p}|_{\mbox{\tiny eq}}
+\delta\eta_{\bf p}\nonumber\\ \\
\kappa_{\bf p}&=&\kappa_{\bf p}|_{\mbox{\tiny eq}}
+\delta\kappa_{\bf p}\nonumber\\\nonumber\\
\langle\phi\rangle&=&\langle\phi\rangle|_{\mbox{\tiny eq}}
+\delta\langle\phi\rangle
\nonumber\\\nonumber\\
\langle\Pi\rangle&=&\langle\Pi\rangle|_{\mbox{\tiny eq}}
+\delta\langle\Pi\rangle\;\;,
\nonumber
\end{eqnarray}
where $\varphi_{\bf k}|_{\mbox{\tiny eq}} \: , \:
\gamma_{\bf k}|_{\mbox{\tiny eq}} \: , \:
\eta_{\bf p}|_{\mbox{\tiny eq}} \: , \:
\kappa_{\bf p}|_{\mbox{\tiny eq}} \: , \:
\langle\phi\rangle|_{\mbox{\tiny eq}}$
and $\langle\Pi\rangle|_{\mbox{\tiny eq}}$
satisfy Eqs.~(\ref{35})-(\ref{40}) and
the deviations $\delta\varphi_{\bf k}$, $\delta\gamma_{\bf k}$,
$\delta\eta_{\bf p}$, $\delta\kappa_{\bf p}$,
$\delta\langle\phi\rangle$ and
$\delta\langle\Pi\rangle$ are assumed to be
small.

Next, we expand Eqs.~(\ref{41a})-(\ref{44}) to
first order in the fluctuations. The following equations result:
\begin{eqnarray}
\label{54}
\delta\dot\varphi_{\bf k}&=&g \: \langle\phi\rangle|_{\mbox{\tiny eq}} \:
\frac{|{\bf k}|}{k_{0}} \:
\delta\gamma_{\bf k}\\\nonumber\\
\label{55}
g \: \langle\phi\rangle|_{\mbox{\tiny eq}} \:
|{\bf k}| \: \delta\dot\gamma_{\bf k}&=&
-4 \: k_{0} \: ({\bf k}^2+\bar m^2) \: \delta\varphi_{\bf k}
-2 \: g \: |{\bf k}| \: k_{0} \:
\delta\langle\phi\rangle\\\nonumber\\
\label{56}
\delta\langle\dot\phi\rangle&=&\frac{4\pi}{(1+Z)} \:
\delta\langle\Pi\rangle\\\nonumber\\
\label{57}
\delta\langle\dot\Pi\rangle&=&-\left(\frac{\mu^2}{4\pi}+
\frac{\delta\mu^2}{4\pi}+\frac{D}{2} \:
G(\mu)\right)
\delta\langle\phi\rangle-\frac{C}{4\pi} \:
\langle\phi\rangle|_{\mbox{\tiny eq}} \:
\delta\langle\phi\rangle\nonumber\\\nonumber\\
&-&\frac{D}{8\pi} \left(\langle\phi\rangle|_{\mbox{\tiny eq}}\right)^2
\delta\langle\phi\rangle+
\frac{4 \: g}{(2\pi)^3}\int_{{\bf k}^{\prime}}
\frac{|{\bf k}^{\prime}|}{{({\bf k}^{\prime}}^2+\bar m^2)^{1/2}}
\delta\varphi_{{\bf k}^{\prime}}\;\;.
\end{eqnarray}

In the RPA regime, the bosonic variables show no dynamical evolution.

To eliminate the quantities $\delta\gamma_{\bf k}$ and $\delta\langle\Pi\rangle$,
can be eliminated by differentiating Eqs.~(\ref{54})~and (\ref{56})
with respect to time, so that Eqs.~(\ref{54})-(\ref{57}) are rewritten
as the second-order differential equations
\begin{eqnarray}
\label{57a}
\delta\ddot\varphi_{\bf k}
&=&-4\bar k_0^{2}\delta\varphi_{\bf k}
   -2g|{\bf k}|\delta\langle\phi\rangle\\\nonumber\\
\label{57b}
(1+Z)\delta\langle\ddot\phi\rangle
&=&-\left( \mu^2 + \Sigma \right)\delta\langle\phi\rangle\
   +16\pi g\int_{\bf {k^{\prime}}} h({\bf {k^{\prime}}})
\delta\varphi_{\bf {k^{\prime}}}\;\;,
\end{eqnarray}
with the notation
\begin{equation}
h({\bf k})=\frac{|{\bf k}|}{\bar k_0^{\mbox{\tiny}}}
\end{equation}
where
\begin{eqnarray}
\label{57c}
\bar k_0&=&\sqrt{{\bf k}^2+\bar m^2}\\\nonumber\\
\label{606d}
\Sigma
&\equiv& \delta\mu^2+\frac{D}{2}4\pi
  G(\mu)+C\langle\phi\rangle|_{\mbox{\tiny eq}}
+\frac{D}{2} \left({\langle\phi\rangle|_{\mbox{\tiny eq}}}\right)^2\;\;.
\end{eqnarray}

The small oscillation-dynamics of the Higgs-Yukawa system is therefore
described by coupled equations of linear oscillators, as usual in the
RPA treatment \cite{KK76}.  In particular, when $g = 0$ these modes
are decoupled and yield two equations describing simple oscillators.

To solve the problem (\ref{57a})-(\ref{57b}) we have to determine
the normal modes of the small oscillations and their frequencies.
Earlier studies have demonstrated that these
elementary excitations can be interpreted as quantum particles.
In our case, $\delta\varphi_{\bf k}$ can be seen as two-fermion
spinless wave function \cite{NTP97}, while $\delta\langle\phi\rangle$ provides
the one-boson physics of the system \cite{KL95}.
The relative momentum of the two-fermion states is $|\bf k|$ \cite{NTP97},
while in the scalar sector, the particles have no momentum dependence.

\section{RPA equations as a scattering problem}
\label{sec:3}
Section~\ref{sec:2} obtained the linear approximation for the Gaussian
equations of motion, Eqs.~(\ref{57a})-(\ref{57b}).
We will now show that these coupled linear oscillator equations
can be analytically solved, to determine the wave functions
and the elementary-excitation spectrum of our system.

We first consider the Fourier transform of the wave functions in the
energy representation, i.~e., the standard relations
\begin{eqnarray}
\label{58}
\delta\varphi_{\bf k}(t)
&=&\int d\omega \: \delta\varphi_{\bf k}(\omega) \:
e^{i\omega t} \nonumber\\ \\
\delta\langle\phi\rangle (t)
&=&\int d\omega \: \delta\langle\phi\rangle(\omega) \:
e^{i\omega t} \;\;, \nonumber
\end{eqnarray}
where $\delta\varphi_{\bf k}(\omega)$ and
$\delta\langle\phi\rangle(\omega)$ are now
energy-dependent amplitudes.

We then substitute Eq.~(\ref{58}) into (\ref{57a})-(\ref{57b}), to
obtain the following equations:
\begin{eqnarray}
\label{66a}
\left(\omega^{2}-4\bar k_0^2\right)\delta\varphi_{\bf k}(\omega)
&=&2g \: |{\bf k}| \: \delta\langle\phi\rangle(\omega) \\\nonumber\\
\label{66b}
\left(\omega^{2}-\mu^2+Z\omega^2-\Sigma\right)\delta\langle\phi\rangle(\omega)
&=&-16\pi g\int_{\bf k^{\prime}}h({\bf {k^{\prime}}}) \:
\delta\varphi_{\bf k^{\prime}}(\omega)\;\;.
\end{eqnarray}

Since the oscillation amplitudes in Eqs.~(\ref{66a}-\ref{66b}) play
the roles of wave functions of quantum particles, it is more
convenient to treat this system as a coupled-channel scattering
problem with appropriate boundary conditions \cite{KL98}.
The following discussion will focus on the scattering process, where the
source is a two-fermion wave. In this case, from Eq.~(\ref{66b}) we
have that
\begin{equation}
\label{66c}
\delta\langle\phi\rangle (\omega)
=\left(\frac{-16\pi g}{\omega^2-\mu^2+Z\omega^2-\Sigma}\right) \:
      \int_{\bf k'} h({\bf k'}) \: \delta\varphi_{\bf k'} (\omega)\;\;.
\end{equation}

Substitution of Eq.~(\ref{66c}) into~(\ref{66a}) then yields the result
\begin{equation}
\label{66e}
\left(\frac{\omega^{2}-4\bar k_0^2}{\bar k_0}\right)
\delta\varphi_{\bf k} (\omega)
=\left(\frac{-32\pi g^2}{\omega^2-\mu^2+Z\omega^2-\Sigma}\right) \:
\frac{|\bf k|}{\bar k_0}
\int_{\bf {k^{\prime}}} \frac{|\bf {k^{\prime}}|}
{\bar k_0^{\prime}}\delta\varphi_{\bf k'}  (\omega)\;\; ,
\end{equation}
where the Green's Function includes the effects of coupling
$\delta\varphi_{\bf k}$
to $\delta\langle\phi\rangle$.

The potential is separable \cite{Ne82}, in the sense that
\begin{equation}
\label{66f}
\langle{\bf k}|V|{\bf k}^{\prime}\rangle=v({\bf k}) \: v({\bf k'})
=\frac{|\bf k|}{\bar k_0} \:
 \frac{|\bf k'|}{\bar k'_0} \;\;.
\end{equation}

In the general solution of Eq.~(\ref{66e}), the two-fermion wave function
$\delta\varphi_{\bf k} (\omega)$ will have two terms. The first one is the
free solution ($g=0$), which represents an incident wave.
The second term is the non-trivial part, arising when $g \neq 0$, which couples
different momenta and is associated with the scattered wave \cite{NTP97,Ne82}.
Therefore,
\begin{eqnarray}
\label{67}
\!\!\!\!
&&\frac{|\bf k|}{\bar k_0} \; \delta\varphi({\bf k},{\bf q};\omega)
=\alpha \; \delta({\bf q}-{\bf k})\;+
\\\nonumber\\
&&\frac{1}{[\omega^2-4\bar k^2_0 + i\epsilon ]} \;
\frac{-32\pi g^2}{[\omega^2-\mu^2+Z\omega^2-\Sigma]} \;
\frac{|\bf k|^2}{\bar k_0}
\int_{\bf k'} \frac{|\bf k|}{\bar k_0} \:
\delta\varphi({\bf k}^{\prime},\bf q;\omega)\;\;,
\nonumber
\end{eqnarray}
where ${\bf q}$ is the relative momentum for two incident
quasi-fermions and $\alpha$ is an overall phase factor.
The outgoing-wave
$(+i\epsilon)$ boundary condition was used to solve Eq.~(\ref{66e}),
but other conditions, e.~g., the incoming-wave condition $(-i\epsilon)$
or Van Kampen wave condition \cite{CTP87}, could alternatively have been chosen.

The integral equation~(\ref{67}) can be solved as usual \cite{Ne82}.
We integrate both sides with respect to ${\bf k}$ to obtain the expression
\begin{equation}
\label{68}
\int_{\bf k} v({\bf k}) \: \delta\varphi({\bf k}, {\bf q};\omega)
=\frac{\alpha}
      {1+\left(\frac{\displaystyle 32\pi g^2}
              {\displaystyle\omega^2-\mu^2+Z\omega^2-\Sigma}\right)
               I^{+}(\omega)} \;\;,
\end{equation}
where
\begin{equation}
\label{68a}
I^{+}(\omega)=\int_{\bf k}
\frac{|{\bf k}|^2}
     {\bar k_0 \; \left[\omega^2-4\bar k_0^2
+ i\epsilon \right]}\;\;
\end{equation}
with $\bar k_0=\sqrt{{\bf k}^2+\bar m^2}\;$, while
$\bar m\equiv m-g\langle\phi\rangle|_{\mbox{\tiny eq}}$
stands for the effective mass of fermion particle and
$\langle\phi\rangle|_{\mbox{\tiny eq}}$ is given by
Eq.~(\ref{40}). Substitution of this last result in Eq.~(\ref{67})
yields the equality
\begin{equation}
\label{69}
\frac{|{\bf k}|}{\bar k_0} \delta\varphi({\bf k}, {\bf  q};\omega)
=\alpha \: \delta({\bf q}-{\bf k})
-\left(\frac{\alpha \bar k_0}{\omega^2-4\bar k_{0}^2+i\epsilon}\right)\;
 \frac{|{\bf k}|}{\bar k_0}
 \: \frac{1}{\Delta^{+}(\omega)} \:
 \frac{|{\bf k}|}{\bar k_0}
\end{equation}
with
\begin{equation}
\label{70}
\Delta^{+}(\omega)
=\frac{1}{32\pi g^2}\left( \omega^2-\mu^2+Z\omega^2-\Sigma \right)
 +I^{+}(\omega)\;\;.
\end{equation}

Finally, substitution of Eq.~(\ref{68}) in~(\ref{66b}) determines the
oscillation frequencies
\begin{equation}
\label{70a}
\omega=2{\bar q_0}^{2}=2\sqrt{{\bf q}^2+\bar m^2}\;\;,
\end{equation}
where $\bf q$ is the relative momentum for two incident quasi-fermions with
mass $\bar m$.

We have therefore found an analytical solution for the elastic channel
of the two-fermion scattering problem defined by Eqs.~(\ref{66a})~and (\ref{66b}).

Thanks to the special form of the interacting potential, the following closed
expression for scattering matrix can also be obtained \cite{Ne82}
\begin{equation}
\label{73}
T({\bf k},{\bf k}';\omega)= v({\bf k})
\frac{1}{\Delta^{+}(\omega)}
v({\bf k}')
\end{equation}
with $\Delta^{+}(\omega)$ given by Eq.~(\ref{70}).

In summary, this section discussed the solutions of the RPA
equations. These elementary excitations describe a coupled channel
scattering problem. The particular case of two-fermion elastic
process was studied.  Given the simple interacting potential, we were
able to obtain closed expressions for the two-fermion wave function and
the scattering matrix. Several dynamical behaviors can be read
off from $\Delta^{+}(\omega)$.  The remaining problem is the divergent
integral $I^{+}(\omega)$ in Eq.~(\ref{70}), to removed with the help of counterterms.

In Sec.~\ref{sec:4} we will see that, in addition to the counterterms
used in the stationary-state calculation \cite{TNa98}, a convenient
wave-function renormalization constant $Z$ will have to be chosen.

\section{Renormalization and bound state solution}
\label{sec:4}
We now use the framework developed in Refs.~\cite{KL95,KL98,NTP97} to
investigate the conditions for the existence of bound states of Dirac
spin-$1/2$ particles in a Higgs-Yukawa system. The standard procedure
is to analyze the positions of the poles of the scattering
matrix~(\ref{73}).  Equation (\ref{70}), however, contains a divergent
integral.  We will next show that the divergent terms can be kept
directly under control with the help of Eqs.~(\ref{94})-(\ref{97}) and a
convenient choice for $Z$, which yields a finite expression for
$\Delta^{+}(\omega)$.

We therefore substitute the counterterms~(\ref{94})-(\ref{97}) in
Eq.~(\ref{70}), and after some algebra, obtain the expression
\begin{equation}
\label{74}
\Delta^{+}(\omega)=\frac{1}{32\pi g^2}[(1+Z)\omega^2-
\mu^2+16\pi g^2G(0)-24\pi g^2M^2L(m)]+I^{+}({\omega})
\end{equation}
with $I^{+}({\omega})$ given by Eq.~(\ref{68a}).

In the interval $0<\omega< 2\bar m$ the integral $I_{\omega}$ is
well defined, and we can let $\epsilon=0$. For $\omega \geq 2\bar m$,
on the other hand, the spectrum defines a continuum.
Straightfoward calculation yields the result
\begin{equation}
\label{VI3}
I(\omega)= Q  - {1\over 8\pi^2}F(\omega)
-\theta(\omega^{2}-4\bar{m}^2){i\over 8\pi}\left[ \omega^2-4\bar
  m^2\right] \; ,
\end{equation}
with
\begin{equation}
\label{VI37}
Q=\frac{1}{4\pi}\left[\Lambda^2
+\left( \frac{\omega^2}{2}-3\bar m^2 \right)\log \frac{2\Lambda}{m}
                \right] \; ,
\end{equation}
where a regularizing momentum cutoff $\Lambda$ was introduced,
and the finite term $F({\omega})$ is given by the relation
\begin{eqnarray}
\label{75a}
F({\omega})&=&
\left(\omega^2-6\bar m^2\right) \log\left(\frac{\bar m}{2m}\right)
+\frac{2(4\bar m^2-\omega^2)^{3/2}}{\omega}
             \tan^{-1}\sqrt{\frac{\omega^2}{4\bar m^2-\omega^2}}
\qquad 0<\omega^2<4\bar m^2 \nonumber\\\\
\label{75b}
F({\omega})&=&
\left(\bar m^2-6\omega^2\right) \log\left(\frac{\bar m}{2m}\right)
+\frac{2(\omega^2-4\bar m^2)^{3/2}}{\omega}
\log\frac{\omega+\sqrt{\omega^2-4\bar m^2}}{\omega-\sqrt{\omega^2-4\bar m^2}}
\qquad \omega^2\geq  4\bar m^2 \;\;. \nonumber\\
\end{eqnarray}

In Eq.~(\ref{VI3}), $\theta$ is the Heaviside function, defined by the relations
\begin{eqnarray}
\label{75bb}
\theta(\omega^{2}-4\bar{m}^2)&=&0 \hskip 1.0cm {\rm{if}} \hskip 1.0cm
\omega^{2}<4\bar{m}^2 \nonumber\\\\
\theta(\omega^{2}-4\bar{m}^2)&=&1 \hskip 1.0cm {\rm{if}} \hskip 1.0cm
\omega^{2}\geq 4\bar{m}^2 \;\;. \nonumber
\end{eqnarray}

\noindent
From Eqs.~(\ref{74})-(\ref{VI37}), we can immediately see that
there is still a logarithmic divergence. To cancel it, we choose the
following wave-function renormalization \cite{AH88}:
\begin{equation}
\label{75c}
Z\equiv 4\pi g^2 L(m) \; .
\end{equation}

The resulting finite expression is

\begin{equation}
\label{76a}
\Delta^{+}(\omega)=-\frac{\pi\mu^2}{g^2m^2}+F(\omega)
-\theta(\omega-4\bar{m}^2){i\over 8\pi}\left[ \omega^2-4\bar
  m^2\right] \, .
\end{equation}

The derivation of Eq.~(\ref{76a}) fixed several counterterms, given by
Eqs.~(\ref{94})-(\ref{97}), and (\ref{75c}), to eliminate the
divergences. These counterterms are not unique, since they are well
defined except for a finite value. This makes the results dependent on
the renormalization scheme. The arbitrary finite constants are usually
determined by high-energy experiments. The dependence on the
renormalization scale is therefore often used to estimate the accuracy
of the theory. In the case of a scalar plasma system, or Higgs-Yukawa
system, Refs.~\cite{TNa98,AH84,AH88} have discussed the determination
of these arbitrary finite constants to obtain the finite stationary
Eqs.~ (\ref{35})-(\ref{40}) in canonical form.

We next face the problem of obtaining the poles of the scattering matrix when
\begin{equation}
\label{76b}
\Delta^{+}(\omega)=0 \; .
\end{equation}

Depending on $\omega$, the system has different dynamical behaviors
\cite{NTP97}.  For $\omega^2<0$, the system is unstable, since the
exponentials on the right-hand sides of Eq.~(\ref{58}) become real.
For $\omega^2>0$, by contrast, the system is in the scattering
regime. The solution of interest lies in the interval of
$0<\omega^2<4\bar m^2$. In this interval the system may have a stable
bound state if there exist $\omega_{_B}$ such that
$\Delta^{+}(\omega_{_B})=0$.  Figure 1 shows $\Delta^{+}(\omega)$ as a
function of $\omega /m$, when $g=1$, for three combinations of $\mu$,
in unit of $m$.  Also for $g=1$, $\Delta^{+}(\omega)$ has a single
(no) zero when $\mu/m<1.794$ ($\mu/m>1.794)$. A natural interpretation
considers that, at fixed coupling, the boson mass determines the range
of the Yukawa potential. When $\mu$ is large, it is more difficult for
the fermions to interact and, consequently, the probability of forming
a bound state, decreases.  This behavior is, however,
compensated by increases in $g$, as shown by Fig.~2, which
plots the condition~(\ref{76b}) in the $(\mu/m\;,\;g)$ plane.

Figure~2 shows that the behavior of $\mu/m \times g$ becomes nearly
linear for $g> 50$. We can apply these results, obtained from the
Higgs-Yukawa model in the RPA mean-field approximation, to recent
observations at the LHC (ATLAS and CMS collaborations), which led to
the announcement of a possible observation of a Higgs boson
\cite{Higgs1}-\cite{Higgs3}. In the experiments five decay channels of
the Higgs boson $\phi$ were observed, i.~e., $\phi \rightarrow \gamma
\gamma$, $\phi \rightarrow b \bar{b}$, $\phi \rightarrow \tau^{+}
\tau^{-}$, $\phi \rightarrow WW$ and $\phi \rightarrow ZZ$. The
production of Higgs bosons in proton-proton collisions is known to
occur through multiple channels, with branching ratios dependent on
the mass of the Higgs boson. Reference~\cite{DataParticle} presents
the branching ratios of the Higgs decay channels as a function of its
mass. For Higgs masses below $130$ GeV/$c^2$, the
Higgs boson is expected to decay mainly in the following fermions: bottom quarks
$b$, charmed quarks $c$, and tau leptons $\tau$.

From Fig.~2 we can determine limits to the intensity of the coupling
constant $g$ of the Higgs-Yukawa model, in the RPA mean-field
approximation, for each decay channel of the Higgs boson. Let us consider a
Higgs boson mass of $m_\phi$=125 GeV/$c^2$, and masses
$m_b$=4.2 GeV/$c^2$, $m_\tau$=1.8 GeV/$c^2$, and $m_c$=1.3 GeV/$c^2$ for the botton
quarks $b$, tau leptons $\tau$, and charmed quarks $c$, respectively
\cite{DataParticle}. We then have the following
ratios between the Higgs-boson mass and the masses of the three fermions:
\begin{equation}
\label{276}
\frac{m_\phi}{m_b} \approx 30 \;\;\; , \;\;\;
\frac{m_\phi}{m_\tau} \approx 70 \;\;\; , \;\;\;
\frac{m_\phi}{m_c} \approx 95 \;.
\end{equation}

Therefore, for these mass ratios and no bound states of fermions
(decay channels), the intensity of the Higgs-Yukawa
coupling for the decay channel $\phi \rightarrow b\bar{b}$ is limited
by the condition $g(\phi,b)<570$.  Similarly, for the decay channel $\phi \rightarrow
\tau^{+} \tau^{-}$ one obtain $g(\phi,\tau)<1300$, and for the decay
channel $\phi \rightarrow c\bar{c}$ one obtains $g(\phi,c)<1800$.

Figure ~2 shows that, in our approach, only two-fermion bound states
can exist for $\mu/m <1.1$.  These results contradict the predictions
of the Higgs boson decaying into top quarks, since $m_t=173 {\rm
  GeV}/c^2$ yields ${m_\phi}/{m_t}={125}/{173} \approx 0.7$. Our
calculations indicate that only Higgs bosons with masses larger than
$190$ GeV/$c^2$ can decay into top quarks. These results are
consistent with the results in the literature \cite{DataParticle}.

Finally, we want to emphasize that the phase diagram in Fig.~2 was
obtained in RPA mean-field approximation.  As discussed in
Refs.~\cite{ETP97,MCNTP83,MCNTP85,CNTP86}, the contribution of the
collisional effects grows with the coupling constant $g$.  For large
$g$ one finds large non-unitary contributions from the collisional
effects. For coupling constants in the range $0 <g <100$, the
corrections, i.~e., the collisional terms, cannot be
neglected. Systematic corrections adding dynamical correlation
effects to the RPA mean-field calculations can in principle be readily
obtained with the help of a projection technique discussed in
Refs.~\cite{TNa98,BFN88,ETP97}.  The resulting occupation numbers
are no longer constant and affect the effective dynamics of the
Gaussian observables.  In particular, a finite matter-density
calculation beyond the mean-field approximation would allow study
of such collisional observables as the transport coefficients.

\section{Conclusions}
\label{sec:5}
Reference~\cite{TNa98} treated the initial-values problem in a
quantum-field theory of interacting fermion-scalar field theories in
the Gaussian approximation.  Although quite general,
the procedure was implemented for the vacuum of an uniform (3+1) dimensional
relativistic quantum Higgs-Yukawa model.  The TDHF
renormalized kinetic equations describing the effective dynamics
of the Gaussian observables in the mean-field approximation were obtained.

The present work has adapted a non-perturbative framework, the Kerman-Lin procedure
[10-11],  to investigate the near equilibrium dynamics close to the
stationary solution of arbitrary interacting fermion-scalar field theories.
As a application, we have chosen to describe the RPA-excitation of the
Higgs-Yukawa system at zero temperature.

We have studied the linearized form of the mean-field kinetic
equations in Ref.~\cite{TNa98} around the stationary (vacuum)
solution.  In this context, the RPA oscillation amplitudes of
excitations were identified with the wave functions of quantum particles
and the resulting equations enabled us to study scattering processes,
non-perturbatively. These RPA equations were solved analytically by
well-known scattering-theory procedures, which yielded a simple form
for the scattering amplitude. We have also shown that the
usual definitions of counterterms can be applied to the resulting
expression, from which relevant physical aspects of the system excitations
can be obtained.  In particular, the results indicate
that bound states exist in certain region of the phase diagram.

We have applied our results to recent observations at the LHC
ATLAS and CMS collaborations. We have obtained limits for the intensity
of the coupling constant $g$ of the Higgs-Yukawa model, in the RPA mean-field
approximation, for three decay channels of the Higgs boson.

Finally, we comment that, in principle, systematic corrections can
readily be added to the RPA mean-field treatment with the help of a
projection technique discussed in Refs.~\cite{TNa98,BFN88,ETP97}.
The no-longer constant occupation numbers
will affect the effective dynamics of the Gaussian observables.
The framework in this paper alse serves as groundwork to discussions
of finite densities and finite temperatures \cite{TTP97}.
In particular, a finite matter-density calculation beyond the mean-field
approximation allows one to study collisional observables, such as transport
coefficients. The extension of this procedure to explore nonuniform
systems is straightforward; unfortunately, it is tedious. In this case, the
spatial dependence of the field are expanded in natural orbitals of
the extended one-body density. These orbitals can be expressed in terms of
a momentum expansion by means of a more
general Bogoliubov transformation \cite{RS80,WP74}.

{\Large {\bf {Acknowledgments}}}

The author P. L. Natti thanks the State University of Londrina for the
financial support received from the FAEPE programs.

\newpage
\centerline{\large\bf Figure Captions}

\vskip 1.0cm

\noindent
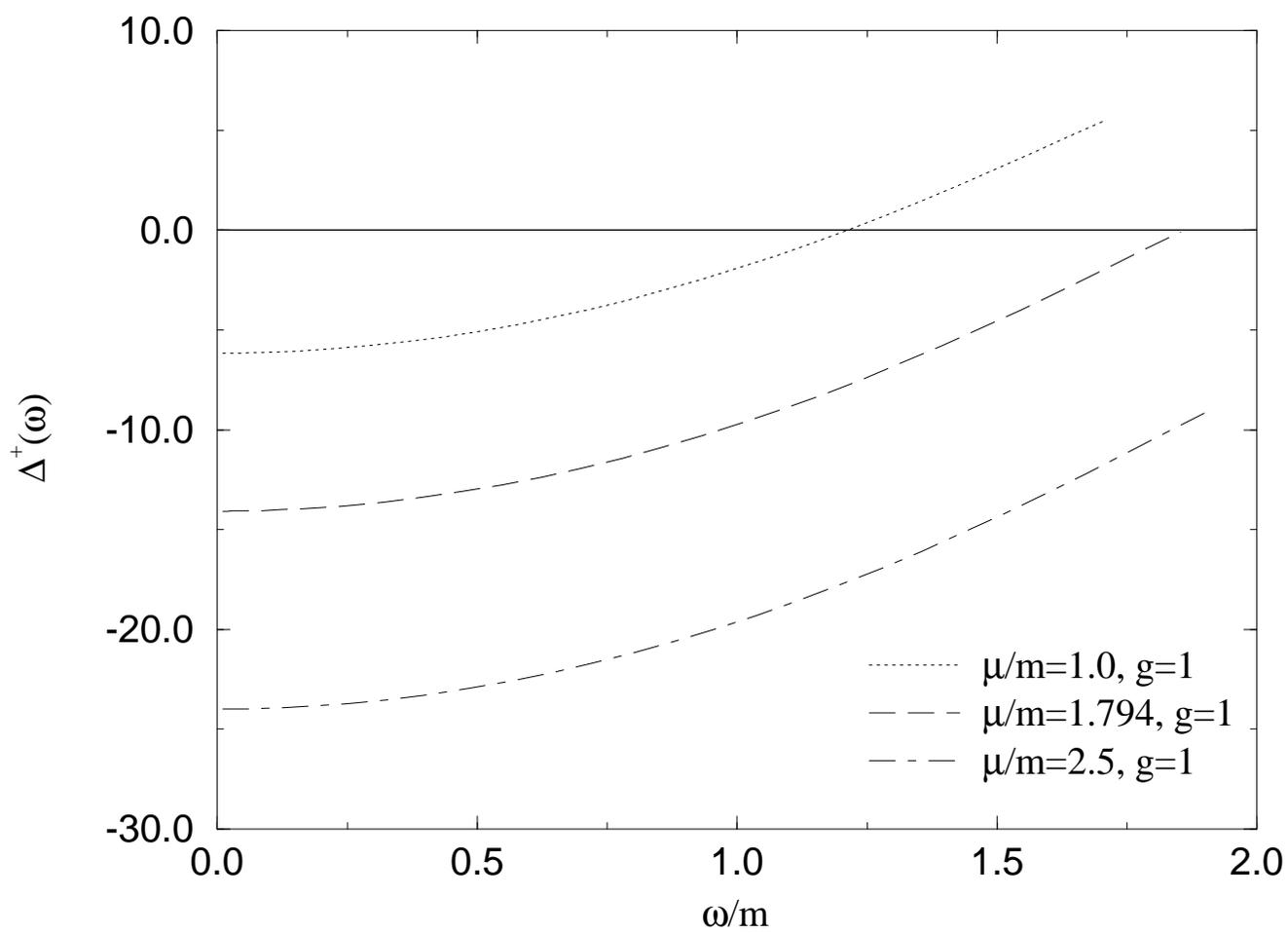
Figure 1. The behavior of the function $\Delta^{+}(\omega)$ as a function
of energy $\omega$, in unit of $m$, for several values of the $\mu/m$, when $g=1$ is fixed.

\vskip 0.3cm

\noindent
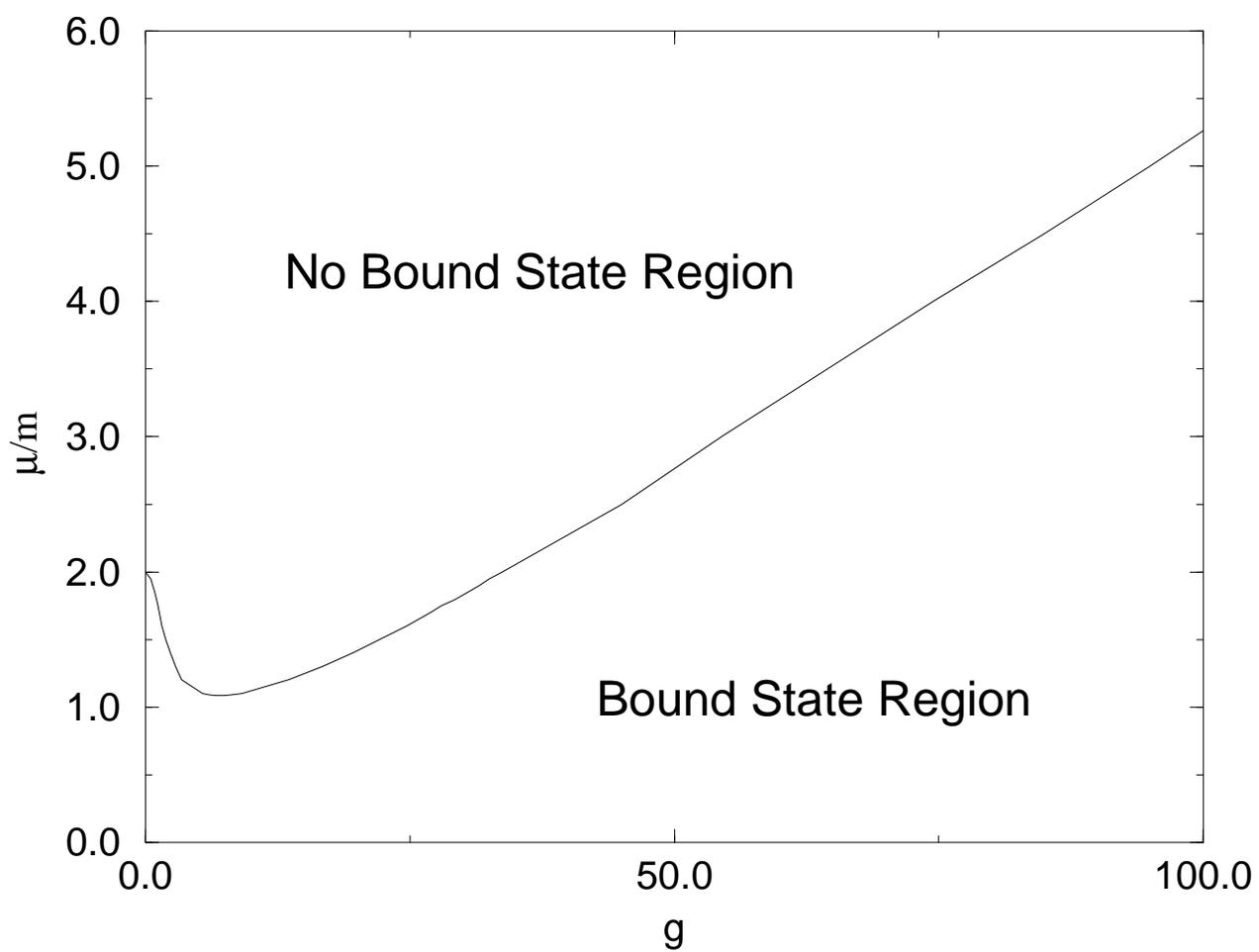
Figure 2. Existence of bound state of two fermion as a function of
parameters $\mu/m$ and $g$.

\end{document}